\documentstyle[aps,prl,epsf,floats]{revtex} 
\begin{document} 
\draft 
\twocolumn[\hsize\textwidth\columnwidth\hsize\csname 
@twocolumnfalse\endcsname 
\title{Frustrated kinetic energy, the optical sum rule, and the 
mechanism of superconductivity} 
\author{Sudip Chakravarty$^1$, Hae-Young Kee$^1$, and Elihu 
Abrahams$^{2}$} 
\address{$^1$Department of Physics and Astronomy, University of 
California 
Los Angeles\\ 
Los Angeles, CA 90095-1547\\ 
$^2$Serin Physics Laboratory, Rutgers University, 
Piscataway, NJ 08854-8019\\} 
\date{\today} 
\maketitle 
\begin{abstract} 
The theory that  the change of 
the electronic  kinetic energy in a direction perpendicular to the 
CuO-planes in high-temperature superconductors is a substantial fraction 
of the condensation energy is examined. It is argued that the consequences 
of this theory based on a rigorous $c$-axis conductivity sum rule are consistent 
with recent optical and penetration depth measurements. 
\end{abstract} 
\pacs{PACS:} 
] 
\paragraph{Introduction.\/} The aim 
of this Letter is to partly resolve a number of 
issues\cite{PWA,Moler,Kirt,Leggett,Basov_Science,Farid,Chak1,VDM,nat} 
related to a 
theory of high-temperature superconductivity known as the interlayer 
tunneling 
theory (ILT) \cite{Thebook} and to propose the efficacy of a 
conductivity sum rule. 
Within a simple version of ILT, one relates the zero-temperature $c$-axis 
penetration depth $\lambda_c$ to the superconducting condensation energy 
\cite{PWA}. 
Here, we point out that the realization of ILT and the interpretation of 
recent measurements 
of $\lambda_c$\cite{Moler,Kirt}, necessarily require more careful analysis 
and that 
the two can be brought into agreement. In addition, we argue that ILT 
accounts for two 
features of $c$-axis optical measurements: (1) the observation that the 
$c$-axis (perpendicular to the 
CuO-planes)  kinetic energy is substantially reduced in the 
superconducting 
state\cite{Basov_Science} and  (2) the 
correlation (``Basov correlation") between 
$\lambda_c$ and the $c$-axis conductivity in the normal state 
\cite{Basov_corr}. 
 
For our purposes, the content of ILT is that a significant portion of 
the superconducting  condensation energy comes from the change in the 
$c$-axis 
kinetic energy as the  electrons enter the 
superconducting state. It is a phenomenological fact that this kinetic 
energy is frustrated in the 
normal state, but that the  frustration is relieved in the 
superconducting state 
as the coherent tunneling of  pairs becomes possible, resulting in a 
sharp 
plasma edge in the reflectivity \cite{Uchida}. 
 
Recently, 
Anderson \cite{PWA} conjectured  that the full 
condensation energy is derived from the 
$c$-axis Josephson energy, which, in turn, determines the penetration depth. 
Then, using 
estimates of the condensation energies, 
he predicted 
$\lambda_c$. On the basis of recent experiments\cite{Moler,Kirt}, it has 
been 
suggested that this prediction is strongly violated  in both  Tl 2201 and Hg 
1201, 
although it appears to hold for LSCO for a 
large 
range of doping. The single-layer superconductors containing one 
CuO-plane per unit cell  are emphasized 
because they pose the most stringent test of ILT. 
 
However, the situation is not so clear. (1) As shown earlier \cite{Chak1}, 
the 
predicted 
$\lambda_c$ should be a factor 2 larger than that predicted in \cite{PWA}. 
(2) The 
measured values of 
$\lambda_c$ are in disagreement. Vortex imaging 
measurements\cite{Moler} give $\lambda_c=17-19\  \mu$m in Tl 
2201, while it is 
$12\  \mu$m in the optical measurements \cite{Basov_Science}, in a similar 
sample. For 
Hg 1201, vortex measurements give $8\  \mu$m \cite{Kirt}, while the optical 
measurements give $6.2\ 
\mu$m, 
again in a similar sample \cite{Basov_comm}, and, disturbingly, 
magnetic measurements yield $1.36\ 
\mu$m\cite{Jcooper}. 
(3) The normal state 
electronic specific heat must be extrapolated to $T=0$ 
from above $T_c$ to determine the condensation energy. 
 
There is an even more fundamental difficulty.  The condensation 
energy is well-defined only within mean field theory. For those 
materials that 
deviate from mean field behavior, that is, those that do not have a 
sharp specific heat 
jump at $T_c$, the condensation energy 
cannot be determined by a simple integration  of the specific 
heat\cite{PWA}. 
The electronic specific heat in Tl 2201 \cite{Loram_Tl} and in Hg 
1201\cite{Melissa_Hg} show large fluctuation effects; we shall show that 
in these cases, agreement with experiments {\em can} be achieved if the 
fluctuation effects are subtracted out. 
 
\paragraph{Sum rule.\/} We now 
discuss the $c$-axis conductivity sum rule \cite{Chak1}. Consider the 
full 
hamiltonian 
$H=H_{\rm rest}+H_c$; the $c$-axis kinetic energy is defined by 
\begin{equation} 
H_c = -\sum_{ij,s}t_{\perp}(ij,l) c^{\dagger}_{il,s}c_{jl+1,s}+{\rm h. 
c.}; 
\end{equation} 
the remainder, $H_{\rm rest}$, contains no interplane interaction 
terms\cite{foot}, but it is 
otherwise arbitrary and may contain impurity interactions  that couple 
to the charge density. 
The hopping matrix element $t_{\perp}(ij,l)$, where  $(i,j)$ refer to 
the sites of 
the two-dimensional 
lattice, and  $l$ to the layer index, can be random in the presence of 
impurities \cite{foot1}. The 
electron operators 
$(c,c^{\dagger})$ are 
also labeled by a spin index 
$s$. We denote the magnitude of $t_{\perp}(ij,l)$ by $t_{\perp}$. 
It is easy to adapt a sum rule derived first by  Kubo \cite{Kubo} 
to get a sum rule 
for the $c$-axis 
optical conductivity 
$\sigma^c(\omega,T)$, which is 
\begin{equation} 
\int_{0}^{\infty}d\omega\ {\rm Re}\ \sigma^c(\omega,T)={\pi e^2 d^2\over 
2\hbar^2 
(Ad)}\langle -H_c\rangle. 
\label{srule} 
\end{equation} 
Here the 
average refers to the quantum statistical average,  $A$ to 
the two-dimensional 
area, and 
$d$ 
 to the separation 
between 
the CuO planes. 
 
The hamiltonian $H_c$  is an effective hamiltonian valid for low 
energy 
processes that do  not involve interband 
transitions. It can be derived by a downfolding 
process, 
 in which all 
the higher energy bands 
are integrated out\cite{Ole}. Because interband processes involve large 
energy 
differences,  a second order 
downfolding procedure is sufficient. 
This is essentially how one derives the effective mass contribution of 
higher 
energy bands. 
 
Since 
$H_c$ is a low-energy effective hamiltonian, the upper 
limit in Eq.~(\ref{srule})  can not exceed 
an interband cutoff $\omega_c$, of order 2-3 eV.  In the 
superconducting state, $\sigma^{cs}(\omega,T)=D_c(T)\delta(\omega)+ 
\sigma^{cs}_{\rm reg}(\omega,T)$, where $D_c(T)$ is the superfluid 
weight. From 
Eq.~(\ref{srule}), it follows that 
\begin{eqnarray} 
D_c(T_1)&=&\int_{0^+}^{\omega_c}d\omega \bigg[{\rm Re}\ 
\sigma^{cn}(\omega,T_2)-{\rm Re}\ \sigma^{cs}_{\rm 
reg}(\omega,T_1)\bigg]\nonumber\\ 
&+&{\pi e^2 d^2\over 2 Ad \hbar^2}\bigg[\langle 
-H_c(T_1)\rangle_s-\langle 
-H_c(T_2)\rangle_n\bigg]. 
\label{srule2} 
\end{eqnarray} 
Here, if $T_2 < T_c$, $\langle -H_c(T_2)\rangle_n$ is to be understood 
as taken in 
the normal state extrapolated to below $T_c$, and $\sigma^{cn}$ is the 
corresponding conductivity. 
 
\paragraph{Lowering of the kinetic energy in the superconducting 
state:\/} 
 
Basov {\em et al.\/}\cite{Basov_Science} have tested the sum rule, 
Eq.~(\ref{srule2}), by setting $T_1\ll T_c$ and 
$T_2=T_c$. 
The result is that up to 0.15 eV, the integral over the conductivity is 
only half the left hand side, so that the 
remaining half must then be the change in the $c$-axis kinetic energy of 
electrons. Since we expect the normal state kinetic energy to 
become less negative as the temperature is lowered \cite{Ando}, this 
measured 
change of the kinetic energy is only a lower bound on the difference in 
the kinetic energy at $T=0$. 
Thus, these experiments support the fundamental statement of ILT that the 
$c$-axis kinetic energy is substantially lowered in the superconducting 
state, 
in contrast to BCS. 
 
\paragraph{The $T=0$ superfluid weight $D_c$ and $\lambda_c$:\/}We relate 
$\lambda_c$ to the change in $c$-axis kinetic energy as 
follows: 
We set 
$T_1=T_2=0$ in Eq.~(\ref{srule2}). 
%%%%%%%%%%%%%%%%%%In optimally doped too? 
From the experiments of Ando {\em et al.\/}\cite{Ando}, it is seen  that 
the $c$-axis response in the 
normal state obtained by destroying superconductivity is insulating as 
$T\to 0$; it follows  that 
$\sigma^{cn}(\omega, T=0)\sim \omega^2$, as $\omega\to 0$. The regular part 
$\sigma^{cs}_{\rm reg}(\omega, T=0)$ is also expected to vanish as a 
power law in a $d$-wave 
superconductor\cite{Bonn}. At high frequencies  the two conductivities 
must, however, approach each other. Consequently it is reasonable to 
hypothesize that the conductivity integral on the right hand side of 
Eq.~(\ref{srule2}) is negligibly small. Therefore, 
\begin{equation} 
D_c(0) = {c^2 \over 8\lambda_c^2} \approx {\pi e^2 d^2\over 2 Ad 
\hbar^2}\bigg[\langle 
-H_c \rangle_s-\langle -H_c \rangle_n\bigg], 
\label{energy} 
\end{equation} 
where we assumed local 
London electrodynamics. 
We emphasize that the choice of the normal state in Eq.~(\ref{energy}) 
is not arbitrary because we have 
assumed that the integral on the right hand side of Eq.~(\ref{srule2}) 
is vanishingly small, and this would 
not be true for an arbitrary state. In any case, the right hand side 
should be a lower bound. 
 
\paragraph{Condensation energy:\/} 
The  attempt to extract the 
condensation energy  from the specific heat data runs into ambiguity, 
except within a mean field treatment. In the presence of fluctuations, 
superconducting correlations, which can primarily be of in-plane origin,
contribute to the energy and significantly to the
specific heat of the
normal state. We suggest that this is indeed the case for Tl 2201 (see below), for 
example. 
To resolve this ambiguity, instead of the conjecture made by Anderson
\cite{PWA}, we propose to subtract the fluctuation effects 
and to use 
the remainder as an effective specific heat from which to 
extract the $c$-axis contribution to the condensation energy.  
The rationale is that free energy can be  decomposed into a singular and a 
non-singular part. The universal singular part is more sensitive to collective 
long-wavelength 
fluctuations, while the non-singular part is dominated by short distance 
microscopic pairing 
correlations. This procedure is well suited to ILT, because, in this theory, 
the  effective ``mean field" condensation energy  can be enhanced due to pair 
tunneling between layers\cite{Chak1}. Note that there is no simple relation between 
$T_c$ and  condensation energy, except in mean field theory.  
 
\begin{figure}[htb] 
\centering 
\epsfxsize=3in 
\epsffile{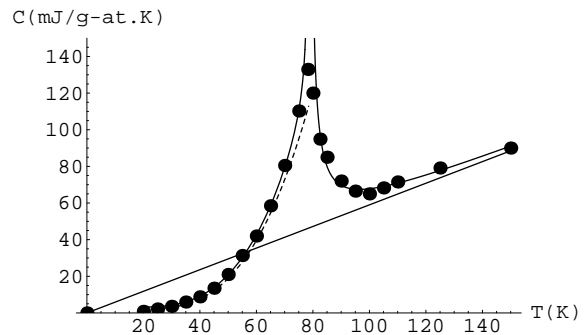} 
\caption{The electronic specific heat data of Tl 2201[15]  fitted to a 
combination of singular and analytic terms (solid line); $T_c =$ 78.7 K. The 
straightline 
is 
$\gamma T$, and the dashed line is the analytic part of the specific heat below 
$T_c$.} 
\label{sp_heat} 
\end{figure}

The fit to the specific heat of Tl 2201 to 
2D Gaussian fluctuation plus non-singular 
terms\cite{Illinois,Gaussian} is shown in Fig. \ref{sp_heat}. 
We have used  
$C(T>T_c)=\gamma T + {g_+/t}$ and 
$C(T<T_c)=c_0 T(1+c_1 t+c_2 t^2)+{g_-/t}$, 
where $t=|1-T/T_c|$ ($\gamma=0.59, g_+=2.38, g_-=0.74, c_0=1.44, c_1=-2.79, 
c_2=2.07$). The fit for 
3D Gaussian fluctuations is considerably worse. This suggests that the in-plane 
contributions dominate.  The data for Hg 1201 are too imprecise to do the 
same, but clearly fluctuation effects are quite prominent and a proper subtraction 
should result in a 
larger prediction for $\lambda_c$. Optimally doped LSCO does not exhibit  
fluctuation effects that are as pronounced. For underdoped  LSCO, we were unable to 
use the  specific heat 
data\cite{Loram_Tl} as they do not seem to fit any simple form. 
 
Condensation energies are obtained from an 
integration of the measured specific heat. In Table \ref{Table} we show 
both results with ($\Delta E_{\rm sub}$) and without ($\Delta E_0$) 
subtracting 
fluctuation effects. By using these values for the right hand side of 
Eq.~(\ref{energy}) we extract the corresponding 
values of $\lambda_c$ as shown along with the experimental values 
\cite{foot3}.  
\begin{table}[htb] 
\begin{tabular}{|c|c|c|c|} 
&LSCO (15\%)&Tl 2201&Hg 1201 \\ \hline 
$\Delta E_0$ &$\sim 150$&$\sim 825$&$\sim 850$\\ 
$\Delta E_{\rm sub}$ &---&$\sim 25$&--- \\ 
$\lambda_{c,0}$ &$\sim 6.5$&$\sim 1.7$&$\sim 2$\\ 
$\lambda_{c,{\rm sub}}$&---&$\sim 10$&---\\ 
$\lambda_{c,{\rm exp}}$&5.5\cite{Uchida2}&$12-19~[2,5]$&$6-8~[3,12]$ 
\end{tabular} 
\caption{Condensation energies (in mJ/g-at) and penetration depths (in 
$\mu$m). Precise error estimates are 
unavailable.} 
\label{Table} 
\end{table} 
The precision of the present Tl 2201 specific heat data is not sufficient to make a 
precise subtraction of the fluctuation contribution. The uncertainty is 
considerable; a reasonable guess for the uncertainty in $\Delta E_{\rm sub}$ 
is $25\pm 15$. Nonetheless, it is clear that using the subtracted value gives a  
much larger penetration depth. 
 
\paragraph{The Basov correlation:\/} 
We 
manipulate the right hand side of 
Eq.~(\ref{srule}) to draw further conclusions.  We perform a canonical 
transformation such that 
$H_c$ is eliminated from the hamiltonian 
$H$. Thus, 
\begin{equation} 
\tilde{H}=e^{-S}He^{S}=H_{\rm rest}+{1\over 
2}[H_c,S]+\cdots,\label{Htilde} 
\end{equation} 
where the antihermitian operator $S$ is defined by $H_c+[H_{\rm 
rest},S]=0$. 
The ground state  $|\tilde{0}\rangle$ of the full hamiltonian $H$ can be 
determined perturbatively in $S$ (or, equivalently $t_{\perp}$) to show 
that the 
ground state expectation value of the 
$H_c$ is given by 
\begin{equation} 
\langle \tilde{0}|H_c|\tilde{0}\rangle=-2\sum_{n\ne 0}\frac{|\langle 
0|H_c|n\rangle|^2}{E_n-E_{0}}+O(t_{\perp}^4), 
\end{equation} 
where $E_n$ and $|n\rangle$ are the eigenvalues and eigenfunctions of 
$H_{\rm rest}$. 
Of course, the same result could be obtained directly without making a 
canonical transformation. We have 
taken this route to hint that the canonical transformation, if carried 
out in infinitesimal steps, could potentially 
be  a powerful method to obtain the effective low-energy 
hamiltonian\cite{Wilson}. 
 
For conserved parallel momentum, the expansion on the right hand side of 
Eq.~(\ref{Htilde}) does not converge in a Fermi liquid  theory because of 
vanishing energy denominators; therefore the expansion would not be 
valid. In  a gapped  state,  
the expansion can be  legitimate because of the absence of vanishing energy 
denominators. 
In a non-Fermi liquid state, the matrix 
elements should vanish  for 
vanishing energy differences, and  the the sum is skewed to 
high energies. Thus,  the energy 
denominator can be approximated by $W$\cite{foot0}, and the sum can 
be collapsed using the completeness 
condition to $\langle \tilde{0}|H_c|\tilde{0}\rangle\approx -\langle 
0|H_c^2|0\rangle/W$. 
The effective hamiltonian $-H_c^2/W$ is identical to the hamiltonian of 
previous realizations of ILT 
\cite{ILT_Science,Chak2}. 
 
Thus $\langle H_c \rangle$ is of order $t_{\perp}^2/W$. 
Then, from Eq.~(\ref{srule}) for example, one can see that on dimensional 
grounds 
the $c$-axis conductivity is 
\begin{equation} 
\sigma_c(T)=a\left({e^2 d t_{\perp}^2\over A W \hbar^2}\right){1\over 
\Omega(T)}, 
\label{conductivity} 
\end{equation} 
where $a$ is a numerical constant weakly dependent on the band 
structure. The inelastic 
scattering rate is proportional to the unknown function $\Omega(T)$. 
Combining the result of the previous paragraph 
with Eqs.~(\ref{energy},\ref{conductivity}), we find 
\begin{equation} 
{c^2\over \lambda_c^2}={4\pi\over a}  \sigma_c(T) \Omega(T) 
\left[u_s-u_n\right], 
\label{penetration} 
\end{equation} 
where $u_{s,n}$ is $\langle 
0|(H_c/t_{\perp})^2|0\rangle_{s,n}$. The average here is 
with respect to the ground state of $H_{\rm rest}$, $|0\rangle$ not 
$|\tilde{0}\rangle$. The product $\sigma_c(T) \Omega(T)$ is independent of 
$T$. 
 
\begin{figure}[htb] 
\centering 
\epsfxsize=3in 
\epsffile{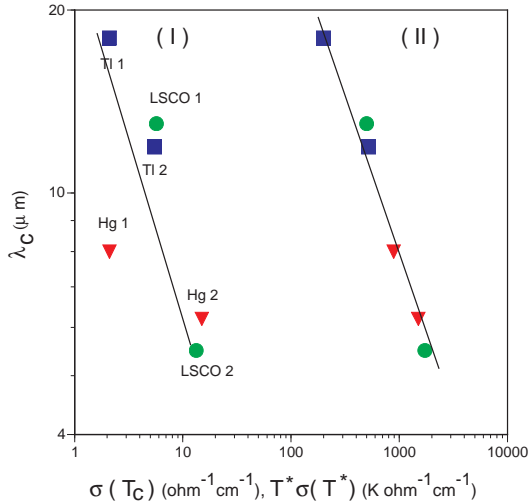} 
\vskip 0.2cm 
\caption{The Basov plot: $\ln \lambda_c$ is plotted against  $\ln 
\sigma_c(T_c)$, as in the original 
Basov plot (I), and against $\ln [T^*\sigma_c(T^*)]$ as discussed here 
(II). The legends in group (II) are the same as those in group (I). Tl1: Ref.~[2]; Tl2: Ref.~[5]; 
Hg1: Ref.~[3], Hg2: Ref.~[13]; LSCO 1 (12\%), LSCO 2 (15\%): Refs.~[15, 
28, 32].} 
\label{Basov_plot} 
\end{figure}

For underdoped to optimally doped materials, the $c$-axis resistivity, 
$\rho_c(T)$, can often be fitted to \cite{foot5} 
\begin{equation} 
\rho_c(T)=b_1 T^{-p}+b_2'T. 
\end{equation} 
The logarithmic behavior\cite{Ando} obtains in 
the limit $p\to 0$. If we express 
Eq.~(\ref{penetration}) in terms of the temperature $T^*$ at which the 
resistivity takes its minimum value, 
we get 
\begin{equation} 
{c^2\over \lambda_c^2}=4\pi \sigma_c(T^*) T^* 
\left\{{b_2(p+1)\over p}\left[u_s-u_n\right]\right\}, 
\end{equation} 
where $b_2=b_2'(d e^2 t_{\perp}^2/\hbar^2 A W)$. 
The expression in the curly brackets depends dominantly on $b_2$, 
which describes the high temperature 
resisitivity. The low temperature behavior enters only through the 
exponent 
$p$. 
Thus, provided the expression in curly brackets is a universal 
constant, a plot of $\ln \lambda_c$ against 
$\ln[\sigma_c(T^*) T^*]$ should be a universal straight line, 
independent of material, with a slope 
of $-1/2$. 
Basov {\em 
et al.}\cite{Basov_corr} discovered a similar correlation by plotting 
$\ln \lambda_c$ against 
$\sigma_c(T_c)$, shown as (I) of Fig.~\ref{Basov_plot}. The correlation 
discussed here, shown as (II), is excellent. 
The data for underdoped LSCO, however, are affected by both the structural 
transition and the (1/8)-anomaly. 
In Fig.~\ref{Basov_plot}, we have taken $T^*\approx T_c$ 
for those optimally doped 
materials that show simply a flattening of $\rho(T)$ close to $T_c$. Thus, 
we see that $b_2$ is indeed inversely proportional to 
$(u_s-u_n)$, which, in ILT, is proportional to the $T=0$ superfluid 
density, 
$n_s(0)$. This can be tested 
further in future experiments\cite{future_work}. 
 
\paragraph{Conclusion:} ILT accounts for a number 
of experimental behaviors, in particular the Basov correlation, and it provides a 
recipe for determining the $c$-axis penetration depth. In  Tl 2201 and Hg 1201 
there must be strong 
superconducting correlations in the normal state. 
The source of these must be both the fluctuation effects not contained 
in the mean field treatment of ILT as well as substantial in-plane pairing 
correlations. Although ILT is not the main driving mechanism for Tl 2201, it may be 
for LSCO and in any case ILT remains an important 
mechanism which can enhance $T_c$ in both single layer and multilayer 
materials\cite{Chak1,Thebook,ILT_Science}. 
 
We thank P. W. Anderson for extensive discussions. The experiments of 
D. Basov, K. A. 
Moler and their coworkers have been instrumental in sharpening our 
thoughts and we thank 
them for many 
discussions. E. A. and S. C. are supported by grants from the 
National Science Foundation. H. -Y. 
Kee is supported by a  Collaborative UC/Los Alamos  Research grant. 
 

\begin{references}
\bibitem{foot}The sum rule remains unchanged if interplane interactions 
that do not transfer electrons 
between the planes are allowed.
\bibitem{foot1}The generalization to further than nearest neighbor 
hopping is contained in the paper by 
Kubo\cite{Kubo}.
\bibitem{foot3}The statement that the experimental penetration depth for 
the optimally doped LSCO 
is $3\mu$m\cite{PWA} is erroneous. The estimates of the condensation energy are 
also incorrect in this 
paper.
\bibitem{foot5}The  conclusion is unchanged if we consider a more 
general expression, 
$\rho_c(T)=b_1T^{-p}+b_2T^q$.
\end{references}
\end{document}